\documentclass[submission,copyright,creativecommons]{eptcs}
\providecommand{\event}{ACL2 Workshop 2017} 
\usepackage{fancyvrb}
\usepackage[shortcuts]{extdash}
\usepackage{textcomp}

\newcommand{\sq}[0]{\textquotesingle}
\newcommand{\bq}[0]{\`{}}
\newcommand{\bqsq}[0]{\bq\sq}

\usepackage{listings}
\title{Hint Orchestration Using ACL2's Simplifier}
\author{Sol Swords
\institute{Centaur Techology, Inc.\\
Austin, TX, USA}
\email{sswords@centtech.com}
}
\def\titlerunning{Hint Orchestration}
\def\authorrunning{S. Swords}
\begin{document}
\maketitle

\begin{abstract}
  This paper describes a strategy for providing hints during an ACL2
  proof, implemented in a utility called \texttt{use-termhint}.
  An extra literal is added to the goal clause and simplified
  along with the rest of the goal until it is stable under
  simplification, after which the simplified literal is examined and a
  hint extracted from it.  This simple technique supports some
  commonly desirable yet elusive features.  It supports providing
  different hints to different cases of a case split, as well as
  binding variables so as to avoid repeating multiply referenced
  subterms.  Since terms used in these hints are simplified in the
  same way as the rest of the goal, this strategy is also more robust
  against changes in the rewriting normal form than hints in which
  terms from the goal are written out explicitly.

\end{abstract}

\section{Introduction}

ACL2's mechanism for giving hints to the prover is very powerful and
general.  Users can provide arbitrary code to evaluate whether a hint
should be given and what that hint should be.  However, it is
surprisingly difficult to accomplish certain things with computed
hints.  We notice that a frequent stumbling block is the difficulty of
predicting the exact form of some term.  A user may know the initial
form of the term but fail to predict its normal form under rewriting;
it may even rewrite to different terms in different cases.

Consider a proof where we want to supply different hints to different cases.
  Suppose the proof splits into two
cases, one assuming \texttt{A} and one assuming \texttt{(not A)},
where \texttt{A} is some term.
Say we want to supply a hint \texttt{H1} for the case where \texttt{A} is assumed true
and \texttt{H2} for the other case.  Users often
provide subgoal hints in such cases, even though these can be broken
either by changes to the relevant functions or by changes to ACL2
system heuristics.  Alternatively, a computed hint can examine the
clause to see when the case split occurs and which case has
resulted---if \texttt{(not A)} is a member of the clause, then
\texttt{A} has been assumed true, and if \texttt{A} is a member of the
clause, it is assumed false:
\begin{Verbatim}[commandchars=\\\{\}]
  :hints ((and (member-equal \sq{}(not A) clause) H1)
          (and (member-equal \sq{}A clause) H2))
\end{Verbatim}
However, a small change in rewriting strategy could cause the exact
form of the term \texttt{A} in the clause to change, in which case
neither hint will fire.

A second kind of problem occurs very frequently when doing inductive
proofs and giving a hint to expand a conclusion term.  The hint may
work in many cases, but frequently there is at least one case where
some variable is substituted for a known value, causing the hint to
fail.  Consider the following example:
\begin{Verbatim}[commandchars=\\\{\}]
(defund add-to-lst (x n)
  (if (atom x)
      x
    (cons (+ n (car x))
          (add-to-lst (cdr x) n))))

(defthm len-of-add-to-lst
  (implies (true-listp x)
           (equal (len (add-to-lst x n))
                  (len x)))
  :hints (("goal" :induct t
           :expand ((add-to-lst x n)))))
\end{Verbatim}
This proof fails because in the base case where \texttt{x} is an atom,
it is known to be \texttt{NIL} due to the \texttt{true-listp} assumption.  This
substitution occurs before \texttt{(add-to-lst x n)} is expanded, and
then the expand hint doesn't match.

We describe a computed hint utility that addresses this problem by
letting ACL2 simplify the terms that will be put into the hint as it
is simplifying the goal itself.  This mechanism allows the user to give different hints
when considering different cases, and the manner of giving these hints
is idiomatic---simply write a term that produces different hints
under different \texttt{if} branches.  It allows terms to be bound to
variables and used multiple times, and since the terms that will
appear in the hints are translated and simplified by ACL2 before the
hint is produced, the user does not need to provide them as translated
terms or in simplified normal form in order to match other occurrences of these terms
in the goal.

In
Section \ref{section:usage} we describe what the utility does and give an
example of how it is used.  In Section \ref{section:design} we
explain the rationale behind certain design decisions, and in
Section \ref{section:sequencing} we describe an extension to the
utility that supports sequencing hints in multiple phases of a proof.  In Section
\ref{section:application} we describe a proof effort in which this
mechanism was used a great deal and give an extended example explaining one theorem proved using the utility.

\section{Related Work}

All major interactive theorem provers allow proofs to be structured at
a high level by proving lemmas that may then be used in later proofs.
They differ more in their approaches to proving individual lemmas.
ACL2 has a robust default proof strategy that the user can modify by
giving hints \cite{acl2:doc}.  HOL and Coq proofs can be constructed
directly at a very low level or may be directed by giving commands
called \textit{tactics} and combining them using tactic combinators
called \textit{tacticals} \cite{hol:description}\cite{coq:reference}.
Isabelle also has tactics, but users more commonly direct proofs by
structuring them using the Isar language, which supports outlining a
proof using a syntax that is reminiscent of mathematical prose, but also
contains directions on how to prove each subgoal
\cite{isabelle_isar:reference}.

The ACL2 Community Books \cite{acl2:doc} contains many libraries that
provide special-purpose computed hints for reasoning in certain
domains.  However, there are few contributions that aid users in
constructing their own hints.  An exception is the community book
\texttt{misc/computed-hint.lisp} written by Jun Sawada
\cite{00-sawada-computed}, which provides a set of practical utilities
intended to improve computed hints.  These support, for example,
pattern matching against terms occurring in the clause while allowing
substitution into the hint to be generated.  We view these utilities
as complementary to the mechanism described here.

\section{Basic Usage}
\label{section:usage}
Our hint utility, \texttt{use-termhint}, is implemented in the
ACL2 community book \texttt{std/util/termhints}.  The utility is
a computed hint form that takes a \textit{hint term} provided by the
user.  Here is the definition of \texttt{use-termhint}:
\begin{Verbatim}[commandchars=\\\{\}]
(defmacro use-termhint (hint-term)
  \bqsq{}(:computed-hint-replacement
     ((and stable-under-simplificationp
           (use-termhint-find-hint clause)))
     :use ((:instance use-termhint-hyp-is-true
            (x ,hint-term)))))
\end{Verbatim}

Initially, this just gives a \texttt{:use} hint and sets up another
hint to fire when the goal is stable under simplification.  The
\texttt{:use} hint instantiates the theorem
\texttt{use-termhint-hyp-is-true}, whose body is
\texttt{(use-termhint-hyp x)}, where \texttt{use-termhint-hyp} is an always-true
function for which no rules are enabled.  The result of the
\texttt{:use} hint is that a hypothesis \texttt{(use-termhint-hyp
  user-hint-term)} is added to the goal clause.  Since no more hints
are given until the goal is stable under simplification, this literal,
including the hint term, is simplified along with the rest of the
clause.

The computed-hint-replacement form produced by the above hint causes a
second computed hint, \texttt{(use-termhint-find-hint clause)}, to
fire when the clause is stable under simplification.  This computed
hint looks in the clause for a hypothesis that is a call of
\texttt{use-termhint-hyp} and extracts a hint from its argument; it
removes the \texttt{use-termhint-hyp} hypothesis using a custom-built clause
processor and then issues the extracted hint.
We describe how the hint is extracted in more detail below.  First,
here is an example of such a computed hint:

\begin{Verbatim}[commandchars=\\\{\}]
 :hints ((use-termhint
          (let* ((f (foo a b))
                 (g (bar f c))
                 (h (baz f d))
                 (i (fa g h)))
             (if (consp g)
                \bqsq{}(:use ((:instance my-lemma
                          (x ,(hq g)) (y ,(hq h)) (z ,(hq i)))))
              \bqsq{}(:expand ((fa ,(hq g) ,(hq h))))))))
\end{Verbatim}

When this hint initially fires, it places the \texttt{let*} term into
the goal as the argument of a new hypothesis \texttt{(use-termhint-hyp ...)}.
ACL2 then beta-reduces and simplifies this term along with the
rest of the clause.  The presence of this additional literal will cause a case split due to its
\texttt{if} test.  When the goal is stable under simplification, this term will
have simplified into something derived from the \texttt{\bqsq{}(:use ...)} subterm
in one case and the \texttt{\bqsq{}(:expand ...)} subterm in the other
case.  For the latter, the resulting term looks like this, if we assume
that none of the \texttt{baz},
\texttt{bar}, or \texttt{foo} subterms were successfully rewritten:

\begin{Verbatim}[commandchars=\\\{\}]
 (CONS
  (QUOTE QUOTE)
  (CONS (CONS (QUOTE :EXPAND)
              ...
              (HQ (BAR (FOO A B) C))
              ...
              (HQ (BAZ (FOO A B) D))
              ...)
        (QUOTE NIL)))
\end{Verbatim}

To extract the intended hint from this term, we use
\texttt{process-termhint}, a simple term interpreter that understands
\texttt{quote}, \texttt{cons}, and \texttt{binary-append} (since this
can be produced when using \texttt{,@} inside backticks).  It also
treats the function \texttt{hq} the same as \texttt{quote}, which we
explain in Section \ref{section:design} below.  This reduces the above
term to the following hint:

\begin{Verbatim}[commandchars=\\\{\}]
 \sq{}(:EXPAND ((FA (BAR (FOO A B) C)
                (BAZ (FOO A B) D))))
\end{Verbatim}

As a special case, if the hint term reduces to \texttt{NIL}, then no
hint is given.

\section{Design Decisions}
\label{section:design}

Readers may note some oddities in the example above:
\begin{itemize}
\item Why \texttt{,(hq g)}, etc., rather than \texttt{,g}?
\item Why use backquote-quote in \texttt{\bqsq{}(:use ...)} rather than just backquote, i.e., \texttt{\bq{}(:use ...)}?
\end{itemize}

\subsection{HQ}
The use of \texttt{hq} distinguishes subterms that should not be
interpreted by \texttt{process-termhint} but simply passed through as
if quoted.  We can't use \texttt{quote} directly because of its
special meaning as a syntax marker rather than a function:
\texttt{,(quote g)} would produce the
symbol \texttt{g} instead of substituting the \texttt{let} binding of
\texttt{g}.  Using \texttt{hq} works because it is not treated
specially by anything other than \texttt{process-termhint}.  (In the logic,
it is simply a unary stub function.)

We could define \texttt{process-termhint} differently to avoid needing to
use \texttt{hq} in most cases: it could treat any term with leading
function symbols other than \texttt{cons} and \texttt{binary-append}
as quotations, in which case we could simply use \texttt{,g} instead
of \texttt{,(hq g)} -- but this wouldn't work if \texttt{g} was bound
to a term that was (or simplified to) a call of \texttt{cons} or
\texttt{binary-append}.  Since we don't wish to require users
to be cognizant of this difference between \texttt{cons} and
\texttt{binary-append} and other function symbols, we decided instead
to require that \texttt{hq} be used to quote terms that
\texttt{process-termhint} should not interpret.

\subsection{Backquote-Quote}

To allow the most general usage of this tool, the hints passed to ACL2
from \texttt{use-termhint} are actually computed hints rather than
literal keyword-value lists.  We can think of
\texttt{process-termhint} as evaluating the hint term
(though it only allows \texttt{cons} and \texttt{binary-append} as
function symbols); however, its result is then passed to
ACL2's computed hint interpreter, which
evaluates it again.  That is why the examples above show hint
keyword/value lists preceded by backquote-quote.  If the quote
occurring immediately after the backquote in the
\texttt{\bqsq{}(:expand ...)}  was omitted, the result from
\texttt{process-termhint} would be \texttt{(:expand ...)} instead of
\texttt{\sq{}(:expand ...)}; the former would cause an error when
evaluated again by ACL2's computed hint mechanism, since it is not a
valid term, whereas the evaluation of the latter yields the expected
keyword/value list.  We actually support this slight abuse as a special case,
adding a quote to any hint that would otherwise begin with a keyword
symbol.  But in the more general case, this double evaluation scheme
allows hint terms to
produce computed hints, as
in the following example:
\begin{Verbatim}[commandchars=\\\{\}]
  :hints ((use-termhint
            (let* ((q (foo a b)))
              \bq{}(my-computed-hint-function
                \sq{},(hq q) clause id stable-under-simplificationp))))
\end{Verbatim}

For the common case where the hint term directly produces a
keyword/value list, we support both the \texttt{\bqsq{}(:key0 val0
  ...)} form, which is doubly evaluated, and the simpler
\texttt{\bq{}(:key0 val0 ...)} form, whose result after evaluation by
\texttt{process-termhint} is quoted so as to
nullify the second evaluation.

\section{Sequencing Hints}
\label{section:sequencing}
It is sometimes useful to provide several stages of hints.  We support
this in the \texttt{use-termhint} utility via a macro
\texttt{(termhint-seq hint-term1 hint-term2)}.  This can be used
inside a term passed to \texttt{use-termhint}.  When simplifying the
initial hint term in which it occurs, \texttt{hint-term1} will get
simplified while \texttt{hint-term2} is wrapped in a call of
\texttt{hide}, which prevents it from being simplified.  Once the
initial simplification is complete, the hint resulting from
\texttt{hint-term1} is applied, and additionally, \texttt{hint-term2}
is provided as the term to a new invocation of \texttt{use-termhint}
with the \texttt{hide} removed.

A simple example:
\begin{Verbatim}[commandchars=\\\{\}]
 :hints ((use-termhint
          (let ((a (bar f c)))
            (termhint-seq
              \sq{}\sq{}(:in-theory (enable my-theory1))
              (if (foo a b)
                  \sq{}\sq{}(:in-theory (enable my-theory2))
                \sq{}\sq{}(:in-theory (enable my-theory3)))))))
\end{Verbatim}

The \texttt{if} test in the second argument to \texttt{termhint-seq}
does not cause a case split until after the first hint (enabling
\texttt{my-theory1}) takes effect, because it is inside a
\texttt{hide}.  The \texttt{let} binding of \texttt{a} does apply to
the occurrence of \texttt{a} in this term.

There is an unfortunate interaction between ACL2's function definition
normalization feature and \texttt{termhint-seq} which that can occur
when \texttt{termhint-seq} is used in a function.  (A user might
create a function rather than putting the whole term in the hint due
to an aesthetic preference to have a theorem appear in its book
without too large a hint list attached.)  Normalization causes
\texttt{if} tests to be pulled out of function calls, even for
\texttt{hide}: \texttt{(hide (if a b c))} becomes \texttt{(if a (hide
  b) (hide c))}.  In the above example, if the hint was defined in a
function rather than given explicitly, the case split would occur
before the first hint was given instead of after.  This can be avoided
by giving the declaration \texttt{(xargs :normalize nil)} when
defining a function that uses \texttt{termhint-seq}.

\section{Application}
\label{section:application}

We used this utility frequently in a proof of the correctness of
Tarjan's strongly connected components algorithm, accessible in the
ACL2 community book \texttt{centaur/misc/tarjan.lisp}.  Much of this
proof involves technical lemmas about the existence of paths through
the graph.  For example, one usage of \texttt{use-termhint} is
in the following theorem:

\begin{Verbatim}[commandchars=\\\{\}]
(defthm reachable-through-unvisited-by-member-cond-2
  (implies (and (tarjan-preorder-member-cond x preorder new-preorder)
                (graph-reachable-through-unvisited-p z y preorder)
                (not (graph-reachable-through-unvisited-p x y preorder)))
           (graph-reachable-through-unvisited-p z y new-preorder))
  :hints ((use-termhint (reachable-through-unvisited-by-member-cond-2-hint
                         x y z preorder new-preorder))))
\end{Verbatim}

We'll informally define enough here to say what the theorem means and
how it is proved.  First,
\texttt{(graph\-/reachable\-/through\-/unvisited\-/p x y preorder)}
says that there exists a path from graph node \texttt{x} to graph node
\texttt{y} that does not include any member of \texttt{preorder}.
Then, \texttt{(tarjan\-/preorder\-/member\-/cond x preorder
  new-preorder)} describes the final visited node set of a
depth-first-search (DFS) when the DFS is run on a node \texttt{x} with
initial visited node set \texttt{preorder}:

\begin{Verbatim}[commandchars=\\\{\}]
(defun-sk tarjan-preorder-member-cond (x preorder new-preorder)
  (forall y
          (iff (member y new-preorder)
               (or (member y preorder)
                   (graph-reachable-through-unvisited-p x y preorder)))))
\end{Verbatim}

I.e., a node \texttt{y} will have been visited by the time the DFS
returns if either it was already visited before the DFS started, or it
can be reached from \texttt{x} without traversing any already-visited
nodes.

The theorem above shows that the visited nodes after a DFS starting
from \texttt{x} don't break any paths to nodes that weren't reachable from \texttt{x}.
That is, given a node \texttt{y} that is not reachable from \texttt{x}
but is reachable from some node \texttt{z}, then it is still reachable
from \texttt{z} when omitting the nodes that are newly visited after a DFS
starting from \texttt{x}.

To prove this, we use the path from \texttt{z} to \texttt{y}.  If that
path is still valid after the DFS (i.e. it doesn't intersect
\texttt{new-preorder}) then we're done.  Otherwise, we'll use that
path to construct a path from \texttt{x} to \texttt{y}.  Since we
assume the path does intersect \texttt{new-preorder}, let \texttt{i}
be a witness to that intersection, i.e. a node in that path that is
also in \texttt{new-preorder}.  The suffix of the path starting at
\texttt{i} is a path from \texttt{i} to \texttt{y} that does not
intersect \texttt{preorder}.  Additionally, since \texttt{i} is not in
\texttt{preorder}, \texttt{tarjan\-/preorder\-/member\-/cond} implies
\texttt{i} is reachable from \texttt{x} without intersecting
\texttt{preorder}.  Composing the path from \texttt{x} to \texttt{i}
with the path from \texttt{i} to \texttt{y} yields a path from
\texttt{x} to \texttt{y} not intersecting \texttt{preorder}, so we
have contradicted our assumption that \texttt{y} is not reachable from
\texttt{x}.

Here is the function producing the hint term used to prove the theorem
via the chain of reasoning described above:

\begin{Verbatim}[commandchars=\\\{\}]
(defun reachable-through-unvisited-by-member-cond-2-hint
  (x y z preorder new-preorder)
  ;; Hyp assumes z reaches y in preorder.  Get the path from z to y:
  (b* ((z-y (graph-reachable-through-unvisited-canonical-witness z y preorder))
       ;; If that doesn't intersect the new preorder, then z reaches y
       ;; via that same path in the new preorder.
       ((unless (intersectp z-y new-preorder))
        \bq{}\sq{}(:use ((:instance graph-reachable-through-unvisited-p-suff
                  (x z) (visited new-preorder)
                  (path ,(hq z-y))))
           :in-theory (disable graph-reachable-through-unvisited-p-suff)))
       ;; Otherwise, get a node that is in both the path and the new-preorder
       (i (intersectp-witness z-y new-preorder))
       ((when (member i preorder))
        ;; this can't be true because it's a member of z-y
        \bq{}\sq{}(:use ((:instance intersectp-member
                  (a ,(hq i))
                  (x ,(hq z-y))
                  (y ,(hq preorder))))
           :in-theory (disable intersectp-member)))
       ;; Since i is a member of new-preorder it's reachable from x
       ((unless (graph-reachable-through-unvisited-p x i preorder))
        \bq{}\sq{}(:use ((:instance tarjan-preorder-member-cond-necc
                  (y ,(hq i))))
           :in-theory (disable tarjan-preorder-member-cond-necc)))
       ;; Construct the path from x to y using the paths from x to i and i to y
       (x-i (graph-reachable-through-unvisited-canonical-witness x i preorder))
       (x-y (append x-i (cdr (member i z-y)))))
    \bq{}\sq{}(:use ((:instance graph-reachable-through-unvisited-p-suff
              (path ,(hq x-y)) (visited preorder)))
       :in-theory (disable graph-reachable-through-unvisited-p-suff))))
\end{Verbatim}

The termhint utility is well suited to this sort of proof because there
are many steps that would be difficult to automate: a rule that would
find a correct witnessing path from \texttt{x} to \texttt{y} in the
example above would need to be either very specific or very smart.
Instead, we guide the proof at a high level by defining the case split
and providing specific hints that resolve each of the cases
separately.

A more common approach to this kind of proof development is to perform
each step as its own lemma.  These lemmas likely aren't suitable as
general rules but can be instantiated to make process in the current
proof.  The theorem discussed above can be proved using four lemmas
corresponding to the cases in the hint function.  This is a reasonable
approach, and having each case stand alone as a lemma might make it
more clear how to debug any problems (though see below for a strategy
to aid debugging proofs that use \texttt{use-termhint}).  However,
stating the lemmas requires repeating the applicable hypotheses and
witness terms for each lemma individually, and this clutter makes it
harder for a human to understand the chain of reasoning.

The main problem in debugging proofs that use \texttt{use-termhint} is
determining which case generated a failed checkpoint.  To remedy this, the book defining
\texttt{use-termhint} provides an always-true function
\texttt{mark-clause} and a corresponding theorem
\texttt{mark-clause-is-true}.  Used as follows, it adds a hypothesis
\texttt{(mark-clause \sq{}my-special-case)}:
\begin{Verbatim}[commandchars=\\\{\}]
  :use ((:instance mark-clause-is-true (x \sq{}my-special-case)))
\end{Verbatim}
Adding such an instantiation to the hints produced by suspect cases
effectively labels each resulting subgoal with a name that
clarifies where it came from.




\section{Conclusion}

The \texttt{use-termhint} utility in some ways goes against the
prevailing philosophy on how to prove theorems using ACL2.  That is,
when possible, it is better to avoid using hints to micromanage the
prover, and instead to create rewriting theories that solve problems more
automatically and robustly.  But sometimes it is necessary to do a
proof using a complicated sequence of reasoning steps that don't seem
to be, in any obvious way, applications of nice rules.  In such cases,
\texttt{use-termhint} provides a robust, convenient, idiomatic method
of structuring a proof at a high level and providing the needed
hints.

\bibliographystyle{eptcs}
\bibliography{../bib}

\begin{thebibliography}{1}
\providecommand{\bibitemdeclare}[2]{}
\providecommand{\surnamestart}{}
\providecommand{\surnameend}{}
\providecommand{\urlprefix}{Available at }
\providecommand{\url}[1]{\texttt{#1}}
\providecommand{\href}[2]{\texttt{#2}}
\providecommand{\urlalt}[2]{\href{#1}{#2}}
\providecommand{\doi}[1]{doi:\urlalt{http://dx.doi.org/#1}{#1}}
\providecommand{\bibinfo}[2]{#2}

\bibitemdeclare{misc}{acl2:doc}
\bibitem{acl2:doc}
\bibinfo{author}{\surnamestart {ACL2 Community}\surnameend}
  (\bibinfo{year}{Accessed: 2018}): \emph{\bibinfo{title}{{ACL2+Books}
  Documentation}}.
\newblock
  \urlprefix\url{http://www.cs.utexas.edu/users/moore/acl2/manuals/current/manual/index.html}.

\bibitemdeclare{manual}{coq:reference}
\bibitem{coq:reference}
\bibinfo{organization}{The Coq Development Team} (\bibinfo{year}{2018}):
  \emph{\bibinfo{title}{The Coq Proof Assistant Reference Manual}},
  \bibinfo{edition}{8.8} edition.
\newblock
  \urlprefix\url{http://coq.inria.fr/distrib/current/refman/index.html}.

\bibitemdeclare{manual}{hol:description}
\bibitem{hol:description}
\bibinfo{author}{Michael \surnamestart Norrish\surnameend} \&
  \bibinfo{author}{Konrad \surnamestart Slind\surnameend}
  (\bibinfo{year}{2018}): \emph{\bibinfo{title}{The HOL System Description}},
  \bibinfo{edition}{3rd} edition.
\newblock \urlprefix\url{http://hol-theorem-prover.org}.

\bibitemdeclare{techreport}{00-sawada-computed}
\bibitem{00-sawada-computed}
\bibinfo{author}{Jun \surnamestart Sawada\surnameend} (\bibinfo{year}{2000}):
  \emph{\bibinfo{title}{{ACL2} Computed Hints: Extension and Practice}}.
\newblock \bibinfo{type}{Technical Report} \bibinfo{number}{TR-00-29},
  \bibinfo{institution}{The University of Texas at Austin, Department of
  Computer Sciences}.
\newblock \bibinfo{note}{ACL2 Workshop 2000 Proceedings, Part A}.

\bibitemdeclare{manual}{isabelle_isar:reference}
\bibitem{isabelle_isar:reference}
\bibinfo{author}{Makarius \surnamestart Wenzel\surnameend}
  (\bibinfo{year}{2017}): \emph{\bibinfo{title}{The Isabelle/Isar Reference
  Manual}}.
\newblock
  \urlprefix\url{http://isabelle.in.tum.de/dist/Isabelle2017/doc/isar-ref.pdf}.

\end{thebibliography}
\end{document}